\documentclass[preprint,aps]{revtex4-1}

\usepackage{amsmath}
\usepackage{subfigure}
\usepackage{epsfig}
\usepackage{pstricks}
\usepackage{overpic}

\begin{document}

\title{Understanding of the Retarded  Oxidation  Effects in  Silicon Nanostructures}

\author{C. Krzeminski}
\affiliation{IEMN-UMR CNRS 8520, D\'epartement ISEN, Avenue poincar\'e, 59650 Villeneuve d'Ascq, France}
\email{christophe.krzeminski@isen.fr}
\author{X.-L. Han}
\affiliation{IEMN-UMR CNRS 8520, D\'epartement ISEN, Avenue poincar\'e, 59650 Villeneuve d'Ascq, France}
\author{G. Larrieu}
\affiliation{LAAS/CNRS, Universit\'e de Toulouse, 7, av. du Colonel Roche 31077 Toulouse Cedex 4, France}
\email{guilhem.larrieu@laas.fr}
\date{\today}

\begin{abstract}
In-depth understanding of the retarded oxidation phenomenon observed during the oxidation of silicon nanostructures is proposed. The wet thermal oxidation  of various  silicon nanostructures such as nanobeams, concave/convex nanorings and nanowires exhibits  an extremely different and complex  behavior. Such effects have been investigated by the mode\-ling of the mechanical stress  generated during the oxidation process  explaining the retarded  regime.  The model describes the oxidation kinetics of silicon nanowires down to a few nanometers while predicting reasonable  and physical stress levels at the Si/SiO$_{2}$ interface by correctly taking into account  the relaxation effects in silicon oxide through plastic flow.
\end{abstract}


\maketitle

\newpage
Retarded oxidation where the oxide growth slows down very rapidly with oxidation duration or with the silicon nanoobject dimension is still a puzzling physical  effect \cite{Okada:1662,liu:1383,Bonafos06}.  This  physical effect can be viewed as a  technological nanoscale tool able to control the nanoobjects  shape,  size distribution  interface  properties  and could be used in many applications \cite{xiaohui11}. However, only very few studies have been dedicated to the understanding of the phenomenon  which remains fragmented and limited  \cite{shir:1333,heidemeyer:4580,Buttner263106}. In this work, oxi\-da\-tion kinetics have been investigated both  on the experimental and theoretical counterparts in order to improve the understanding of the mechanisms of retarded mechanisms and to quantify the amount of stress  generated  at the Si/SiO$_{2}$ interface in silicon nanostructures.

With the current top-down fabrication capabilities, etched  silicon nanostructures including nanobeams, nanorings and nanowires have  been fabricated with a high resolution \cite{Han20112622} and then  wet oxidized at 850$^{\circ}$C. The seminal work of Kao {\it et al.} \cite{Kao:1486748,Kao:2412}  with micrometer size of 2D cylindric  object is revisited but in the nanometric range. Experimentally, the oxidation kinetics have been observed to be strongly dependent on the size and the geometry of the nanoobject. Fig. \ref{fig:fig1}.a) summarizes the evolution of the oxide thickness as a function of the oxidation duration in the case of convex (SiNWs) and concave (Si nanorings) structures. The oxide growth rate is strongly limited with the oxidation time but is faster in a convex structure than in a concave one. The influence of  the geometrical effect is stronger with smaller  inner radius (i.e. 70 nm compared to 430 nm). For the convex case, a higher oxide growth rate is related to the larger radii.  Then, in order to investigate  experimentally the influence of silicon nanostructure dimension,  nanobeams and nanowires  of height 240 nm have been oxidized  for 10 and 20 minutes. As shown in insert \ref{fig:fig1}.b).(1), a non-uniform oxide growth is  classically observed along the sidewall of the beam due to the great influence of  the top and bottom corners corresponding to a convex and concave structure respectively. An oxidized one-dimensional nanostructure  with diameters from 40 nm to 140 nm  demonstrated completely different shapes as shown in  insert \ref{fig:fig1}.b).(2) with the presence of a pinching effect at the bottom of Si nanobeam structures. In order to compare the oxidation between the two structures, the oxide thicknesses have been measured in the middle  of these vertical structures described in SEM images and are plotted  in Fig. \ref{fig:fig1}.b). The oxide growth on Si nanobeams of previous width L is  clearly thicker than a SiNWs with the diameter $d=L$. A size dependent oxidation kinetic was not observed in these structures whatever the nanobeam width considered.  These experimental results  illustrate that the silicon oxidation retarded mechanism is strongly dependent at the nanoscale level on the nanoobject  i) dimension ii) size  and iii) shape.

These dependences cannot be explained by the standard Deal and Grove  oxidation model \cite{deal:3770}, as for example,  a larger oxidant concentration for the smallest particles should in principle lead to a higher oxidation rate. Two main theories have been put forward to explain the retarded/self-limiting kinetics factor. The first one is the ``stress limited reaction rate'' assumption \cite{heidemeyer:4580}  with a  radial stress build-up at the Si/SiO$_{2}$ interface  assumed to be linearly dependent of the oxidation time  up to a critical stress estimated to a few GPa  where the oxidation rate would be completely negligible.  The second theory is the ``diffusion limited mechanism'' associated to a significant increase  in the activation energy of the oxidant diffusivity in the highly stressed region is put forward \cite{liu:1383}.  In this framework, the  origin of self-limited effects  would be the  oxidant species supply  at the  interface.  Despite the fact that an unknown and uncontrolled amount of strain is introduced, no  quantitative determination of the  mechanical stress  build-up during processing is provided by the two approaches.  
 
In order to model the oxidation of cylinder nanoobjects, the extended Deal and Grove model  in cylindrical coordinates has been used \cite{Kao:2412}. The  wet oxidation rate  $v$  at the Si/SiO$_{2}$ interface is given by  (Eq. \ref{eq:eq_one}) :
\begin{equation}
v=\frac{(\alpha-1)C^{*}}{N}\cdot \frac{1}{\displaystyle \frac{1}{k_{Si}^{\sigma}}\pm \displaystyle \frac{a}{D^{p}} \log   ( \frac{b}{a}  )}
\label{eq:eq_one}
\end{equation}

with  respectively $a$ ($b$) the inner (outer) radius, $\alpha$  the volume expansion factor of silicon to oxide conversion (2.25),  $N$  the number of oxidant molecules incorporated into a unit volume of silicon oxide, the $+$ (resp. -) sign denotes the convex (concave) surface. This equation  classically takes into account that the surface curvature influences the oxidant concentration and in the convex (concave) configuration, the   concentration increases (decreases).  In our approach, both the reaction rate k$_{Si}^{\sigma}$ at the Si/SiO$_{2}$  interface and the diffusivity in the silicon oxide $D_{SiO_{2}}$  are stress dependent. The reaction rate $\displaystyle k_{Si}^{\sigma}$ is directly proportional to  the linear rate constant $(B/A)_{[110]}(T)$ (5.18 10$^{-09}$ nm/s at  850$^{\circ}$C)  defined in the Deal and Grove approach by introducing  $C^{*}$  the oxidant solubility in the silicon dioxide and is strongly dependent on the radial stress component $\sigma_{r}$ at the Si/SiO$_{2}$ interface : 

\begin{equation}
\displaystyle k_{Si}^{\sigma}= \frac{N}{C^{*}}\cdot  (B/A)_{[110]}(T) \exp \big (\frac{ \sigma_{r} V_{k}}{k_{B}T} \big )\\
\label{eq:eq_two}
\end{equation}

where  $k_{B}$ is the Boltzmann constant, $T$ is the oxidation temperature and V$_{k}$ (15 \AA$^{3}$)   corresponds  to the  activation volume. A compressive  radial stress ($\sigma_{r}<0$)  slows down the linear rate oxidation rate. The  term $(B/A)_{[110]}(T)$  takes into account the influence of the $[110]$ crystalline orientation and the  factor taking into account orientation effects has been calibrated with planar bulk oxidation experiments. Next the oxidant diffusivity, D$_{SiO_{2}}$ :

\begin{equation}
D_{SiO_{2}}=\frac{N}{2 C^{*}}\cdot B(T) = \frac{N}{2 C^{*}}\cdot B_{0}(T) \cdot  \exp \big (\frac{-P V_{d}}{k_{B}T}\big )
\label{eq:eq_three}
\end{equation}

is linearly dependent on the initial parabolic constant $B_{0}(T)$ (2.68 10$^{-13}$ nm$^{2}$/s at 850$^{\circ}$C)  and is limited   by  a compressive ($P>0$)   hydrostatic pressure  $P=-0.5\cdot(\sigma_{r}+\sigma_{\theta})$  in the silicon oxide  (V$_{d}$=45 \AA$^{3}$). These assumptions are often estimated  to be equivalent to a diffusivity dependence with oxide density \cite{Sutardja89}.\\  

A major issue in oxidation modelling is a proper description of the mechanical behavior of silicon dioxide as shown, in Fig. \ref{fig:fig2}, and its ability to store or to dissipate mechanical energy. A shortcoming  is also observed for the viscous standard approach \cite{Kao:2412,Senez94}   since the compressive radial stress at the interface is  inversely proportional to the curvature  radius of the nanoobject and strongly overestimates the stress level \cite{Rafferty89a}.  The main reason is that the  irreversible atomic rearrangements occurring with large shearing forces  \cite{Falk2010} are neglected.  This plastic flow can been  described by a  shear dependent  viscosity \cite{eyring:283} :

\begin{equation}
\eta(\tau)=\eta_{0}(T) \frac{\tau / \sigma_{c}}{\sinh(\tau/\sigma_{c})}
\label{eq:eq_four}
\end{equation}

where $\eta_{0}(T)$ is the low stress viscosity, $\tau$ is  the critical resolved  shear stress and  $\sigma_{c}$ is the critical stress threshold where plasticity flow should appear (1 GPa). The low  stress viscosity value  (1.4 10$^{18}$ Poisse at 850$^{\circ}$C) considered is characteristic of a wet oxide with a high viscosity induced by the presence of  hydroxyl content \cite{Hetherington64}.  Following the expression of the  critical resolved shear stress $\tau(r)= \frac{2 \eta a \nu}{r^{2}}$, it can be underlined that the  oxidation growth rate (Eq. \ref{eq:eq_one}), the shear dependent viscosity (Eq. \ref{eq:eq_two})  and finally the critical shear stress   are coupled to each others. The fact that all of these equations   must be  self-consistently solved is often overlooked or not exactly taken into account. Following Rafferty {\it et al.} \cite{Rafferty89b},  the radial ($\sigma_{r}$) and   tangential   ($\sigma_{\theta}$) stress field  component in the silicon dioxide of a cylinder structure  (see Fig. \ref{fig:fig1}.b)  can be expressed  as :

\begin{equation}
\left  \{
\begin{array}{l}
\sigma_{r}(r)=\pm \frac{1}{2} \sigma_{c} \big [ \big ( \displaystyle \ln \frac{R^{2}}{b^{2}}\big )^{2}  -\big ( \ln \frac{R^{2}}{r^{2}} \big )^{2} \Big ] \\
\displaystyle \sigma_{\theta}(r)=\sigma_{r}(r)-2\tau(r)\\
\end{array}
\right.\\
\label{eq:eq_six}
\end{equation}

with the reduced parameter $R=\sqrt\frac{\displaystyle 4\eta_{0}av}{\sigma_{c}}$. Compared to a standard viscous approach \cite{Kao:2412} with a constant viscosity,  the radial stress build-up has a logarithmic dependence on the curvature radius  which gives us the opportunity to model the oxidation of cylinder shape nanostructures.\\

Fig. \ref{fig:fig3}.a) shows that  the influence of  the  concave or  convex character on the  oxidation kinetics can be well  predicted by the  model. As shown in Fig. \ref{fig:fig3}.b) a  substantial  non-linear increase  is observed for the compressive radial stress component at the Si/SiO$_{2}$ interface up to a few GPa, coupled with an initial tensile hydrostatic pressure  (inset Fig. \ref{fig:fig3}.b).   The  radial stress build-up is not linear with time as assumed in a previous study \cite{heidemeyer:4580}.  These elements   clearly indicate that  a  reaction limited process takes place in the convex configuration. The situation is totally different in the concave case where a radial compressive stress build-up  remains limited whereas the compressive hydrostatic pressure clearly impacts the oxidant diffusivity. In that case, the major limiting  factor is the diffusion mechanism which reduces the  oxidant supply. A quasi self-limited oxidation for  70 nm concave structure is observed and can  be correlated to  the occurrence of both a   diffusion and reaction limited regime. As summarized by table \ref{tab:table1}, the dominant  retardation mechanism is strongly dependent on  the surface shape but can be explained by the variation of the stress field component at the Si/SiO$_{2}$ in agreement with previous results \cite{Kao:2412}.

Fig. \ref{fig:fig4}.a) presents the linear rate modeling with  oxidation duration and SiNW diameters reported in Fig \ref{fig:fig1}.b). A strong decrease in the reaction rate  with radial stress build-up as a function of SiNW diameters is predicted which could be  directly correlated to the  large radial stress build up depicted in  Fig. \ref{fig:fig4}.c). Large non-linear build-up as a function of the oxidation time and SiNW diameters for the compressive radial stress down to 4 GPa which remains compatible with  interfacial stress estimated   using contact-resonance atomic force microscopy  \cite{stan2010}.  This effect causes the initial retarded effects observed in the oxidation of convex nanostructures. These simulations explain well the difference in terms of behavior  between the nanobeams and the nanowires since finite elements simulations estimate a much lower  compressive stress build-up in the nanobeam. On the other hand,  Fig. \ref{fig:fig4}.b) presents the evolution of the parabolic rate with SiNW diameters. A decrease in the parabolic rate  is observed after a significant delay  which can be correlated to the tangential  stress  relaxation as shown in inset of Fig. \ref{fig:fig4}.c). This  diffusion limited effect generated by a compressive hydrostatic pressure  is  probably  much more difficult to control  as a time and diameter dependence is observed in Fig.  \ref{fig:fig4}.b).

In summary, retarded oxidation kinetics have been investigated at the nano\-scale level for different silicon nanoobjects. We have demonstrated the influence of the  nanoobject dimension, size and shape on the oxidation behavior. All of these effects have  been correlated to the  interfacial stress build-up during oxidation. Given the  deformation rate level, modelling aspects show that  plastic relaxation needs to be considered  in order to estimate i) a physical mechanical stress build-up at the interface and ii) the interface velocity.  Both reaction or diffusion  limited mechanisms must be considered to desccribe retarded oxidation effects in silicon nanostructures.

\section*{Acknowledgement}
This work was  partially supported by the European Commission through the NANOSIL Network of Excellence (FP7-IST-216171) and the RTB platform (French national nanofabrication network).

\newpage
\begin{table}[]
\caption{\label{table:cesl_table} Main physical mechanism governing the retarded or self-limited oxidation with decreasing  convex or concave nanobjects size.}
\begin{tabular}{lccl}
Character&$\sigma_{r}$&$P$&Origin\\
\hline\hline
Convex&$\nearrow$&$\ll$&Limited reaction rate\\
Concave&$\ll$&$\nearrow$&Limited diffusion mechanism\\
\hline\hline
\label{tab:table1}
\end{tabular}
\end{table}

\newpage
\noindent{FIGURES CAPTION}

\begin{figure}
\center
\subfigure[]{\begin{overpic}[width=6.5cm]{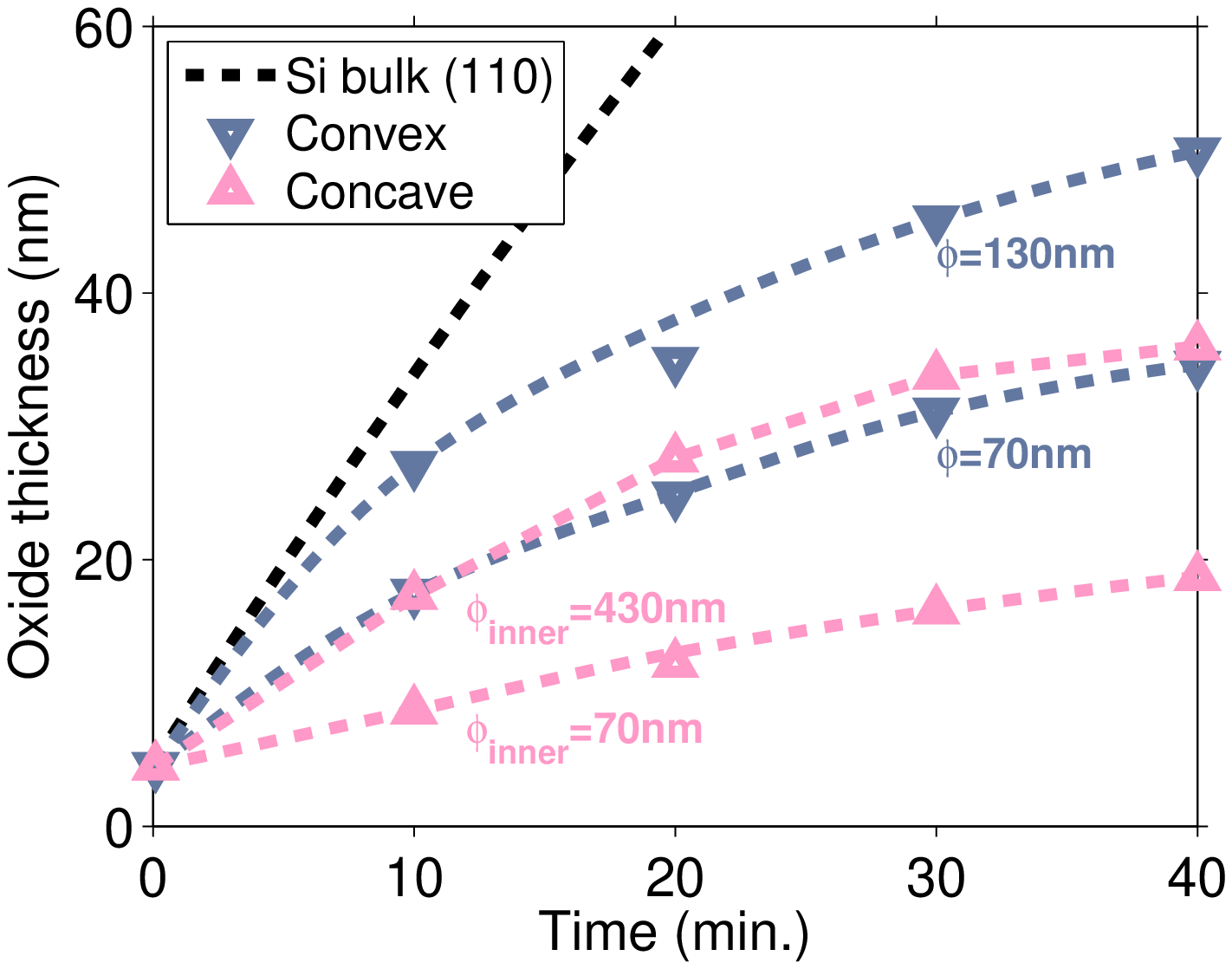}
             \put(60,10){\includegraphics[width=2cm]{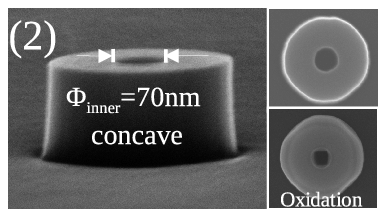}}
             \put(46,52){\includegraphics[width=2cm]{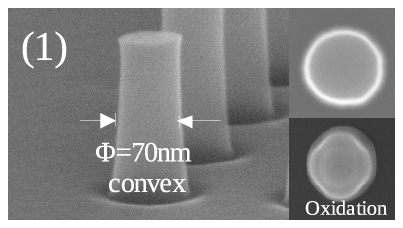}}
             \end{overpic}
             }
\subfigure[]{
            \begin{overpic}[width=6.5cm,angle=0]{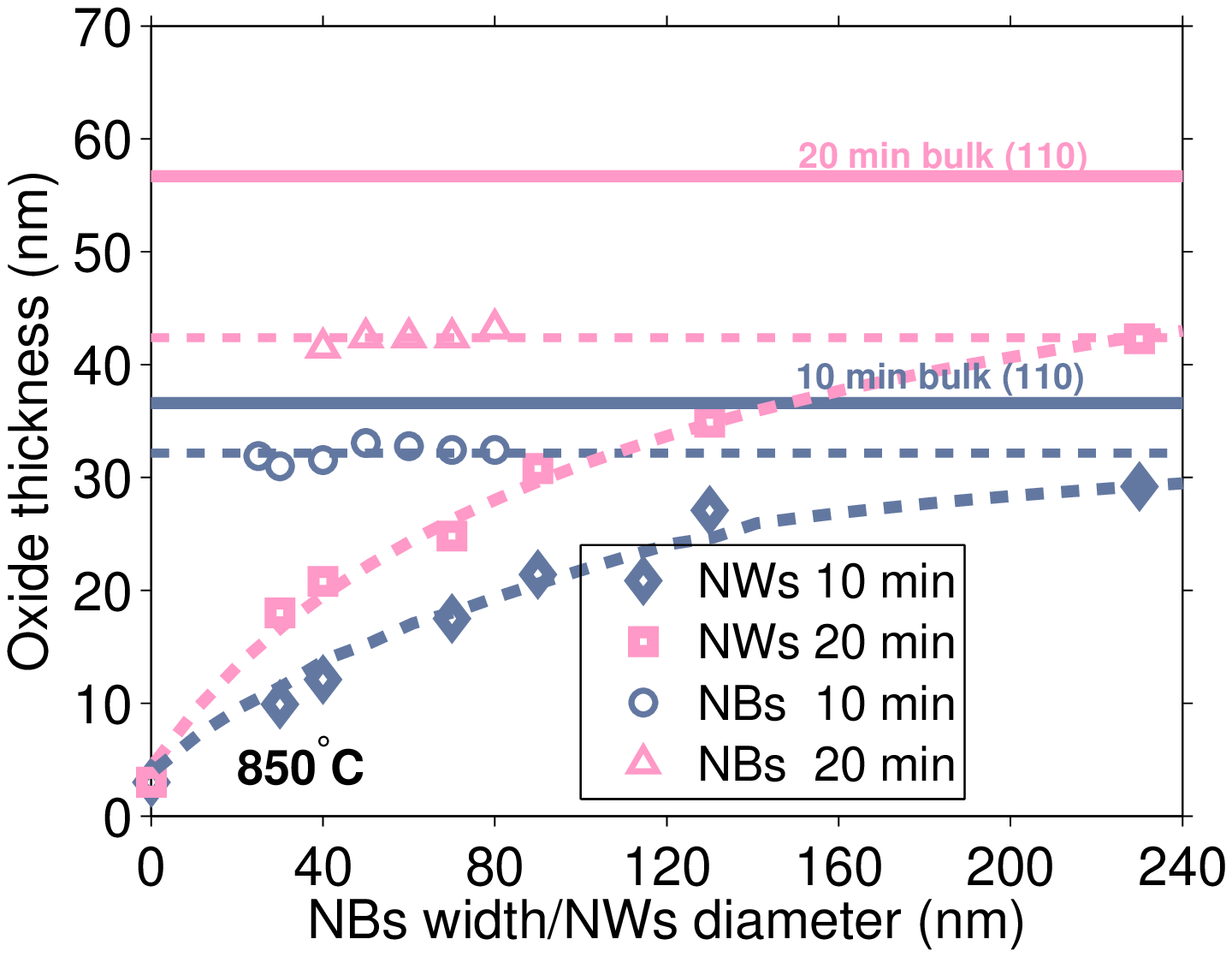}
            \put(14,49){\includegraphics[width=0.95cm]{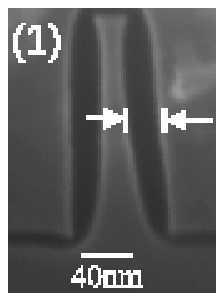}}
            \put(75.5,11.5){\includegraphics[width=0.95cm]{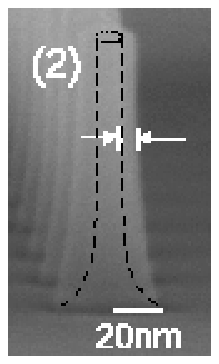}}
            \end{overpic}
            }
\caption{\label{fig:fig1}}
\end{figure}

Fig. \ref{fig:fig1}: (a) Convexity vs concavity: the various symbols present the experimental oxide thickness for SiNWs (convex nanostructure) with diameters 70 nm, 130 nm  and silicon nanorings (concave nanostructure) with inner diameters of 70 nm and 430 nm  as a  function of the oxidation time (b) Oxide thickness  in the middle edge of  the Si nanobeams and SiNWs  for 10 min and 20 min at 850$^{\circ}$C. The experimental trend is described by the dashed lines. Inset SEM images: (1) the Si nanobeams (cross-section view), (2) SiNWs (titled view) after oxidation  where the position set to measure the oxide thickness is reported.

\begin{figure}
\center
\includegraphics[width=6cm,angle=0]{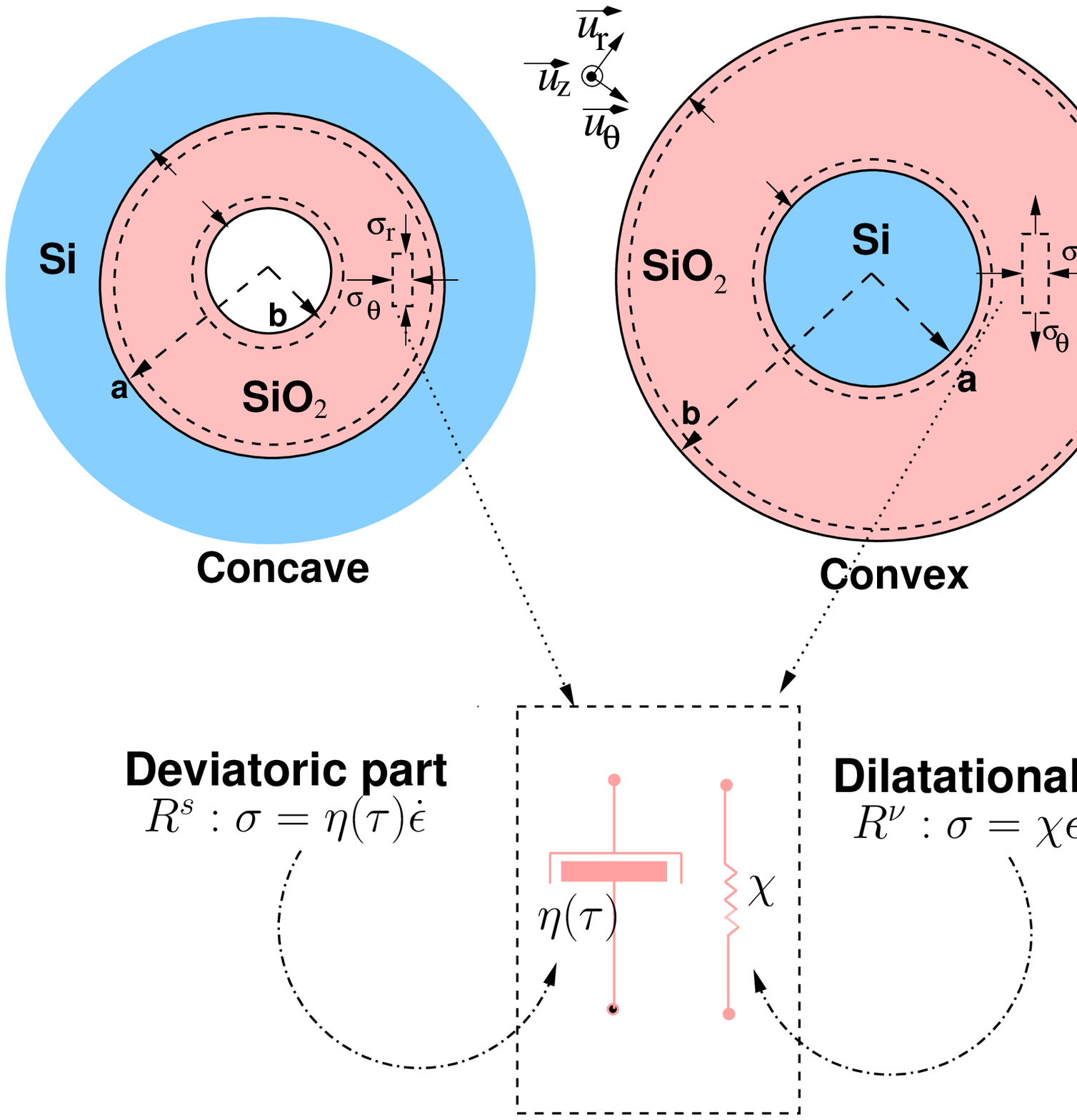}
\caption{\label{fig:fig2}}
\end{figure}

Fig. \ref{fig:fig2}: Schematics of the concave and convex cylinder nanostructure oxidation and  resulting stress field in the silicon oxide. The strain in the oxide could be divided into two components the deviatoric part associated to shape modification and the  dilatational part often neglected.  Plasticity effects are introduced by considering a non-linear shear dependent viscosity $\eta (\tau)$.

\begin{figure}
\subfigure[]{\includegraphics[width=6cm]{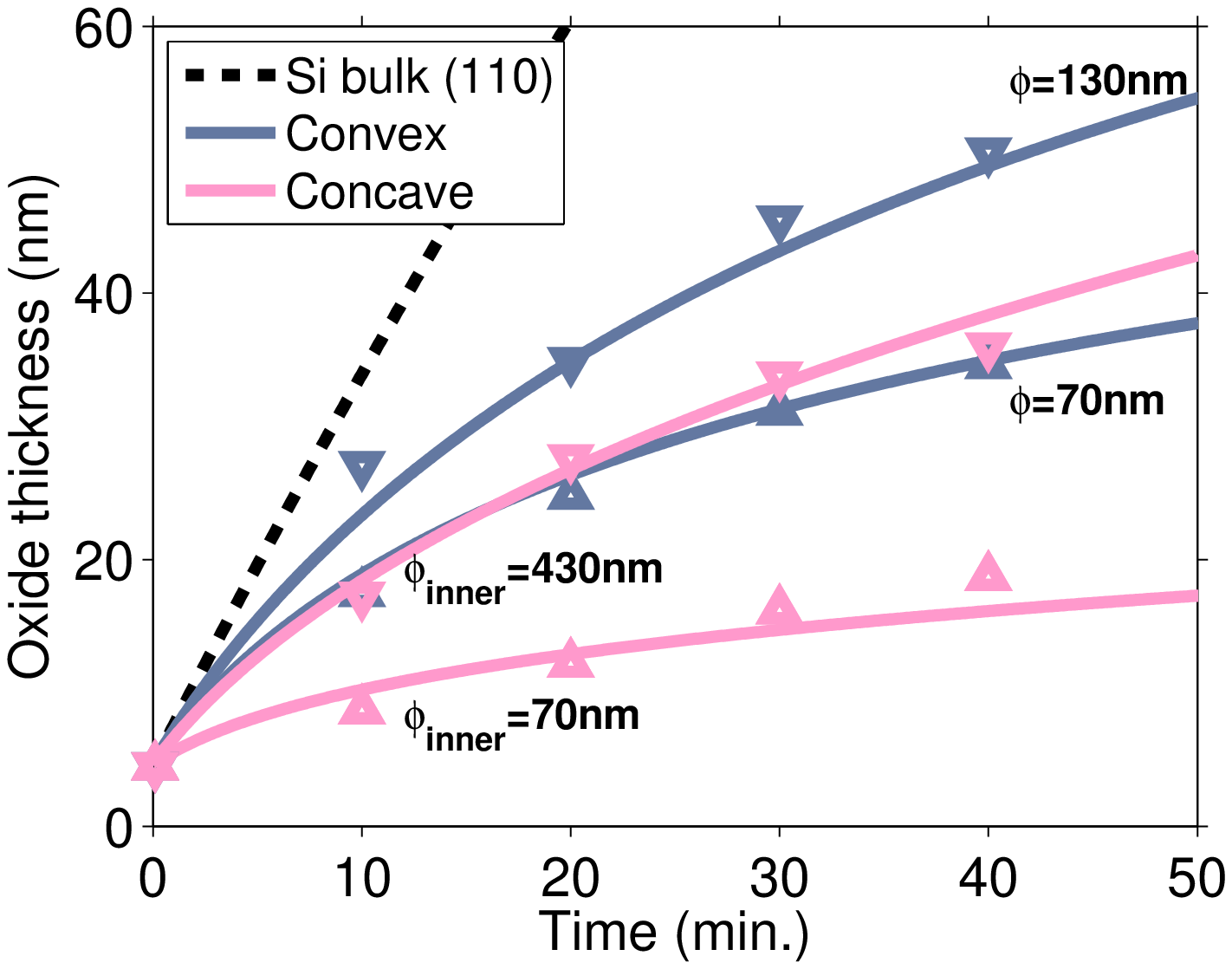}}
\subfigure[]{\begin{overpic} [width=6cm]{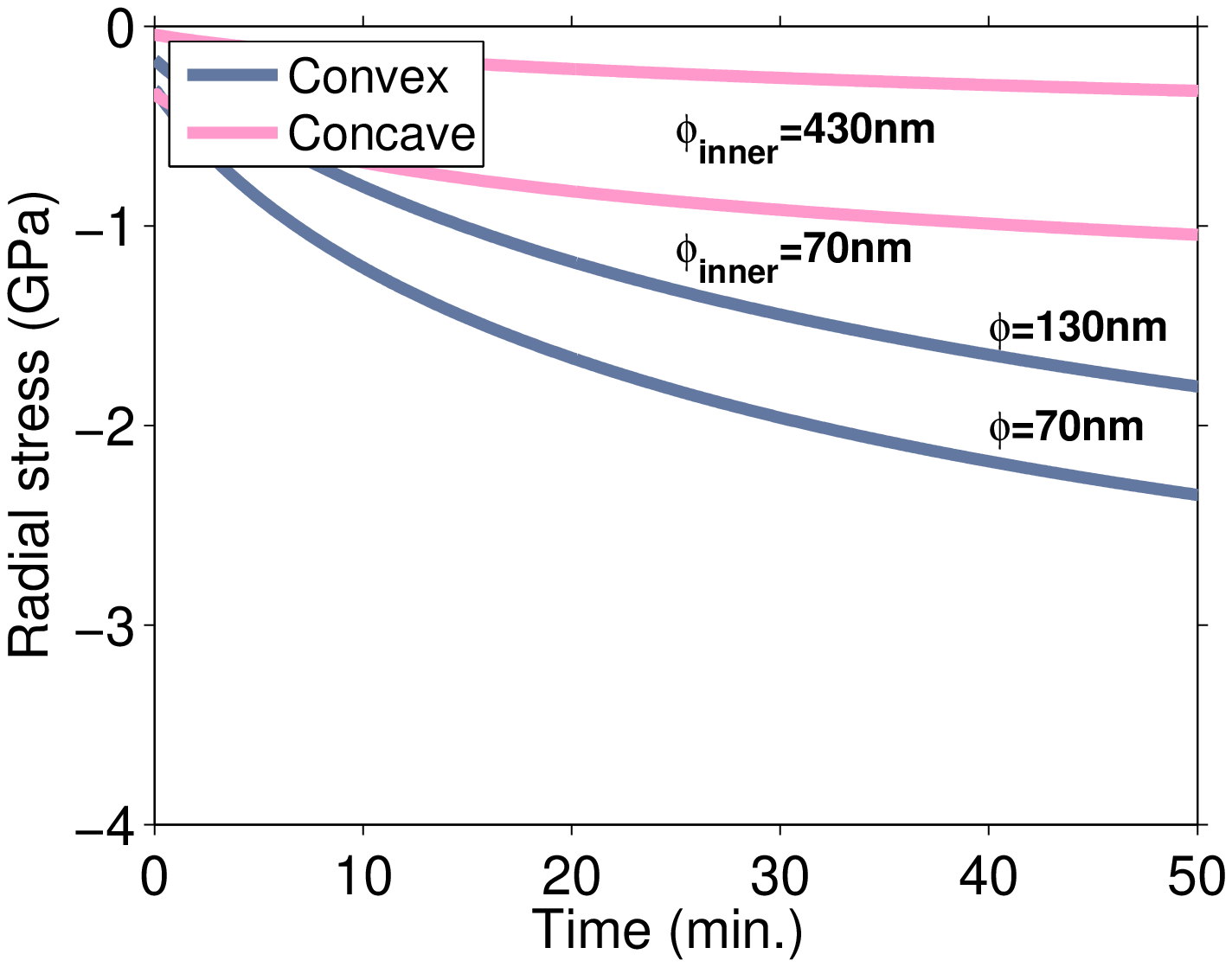}
    \put(14,11){\includegraphics[width=2.8cm]{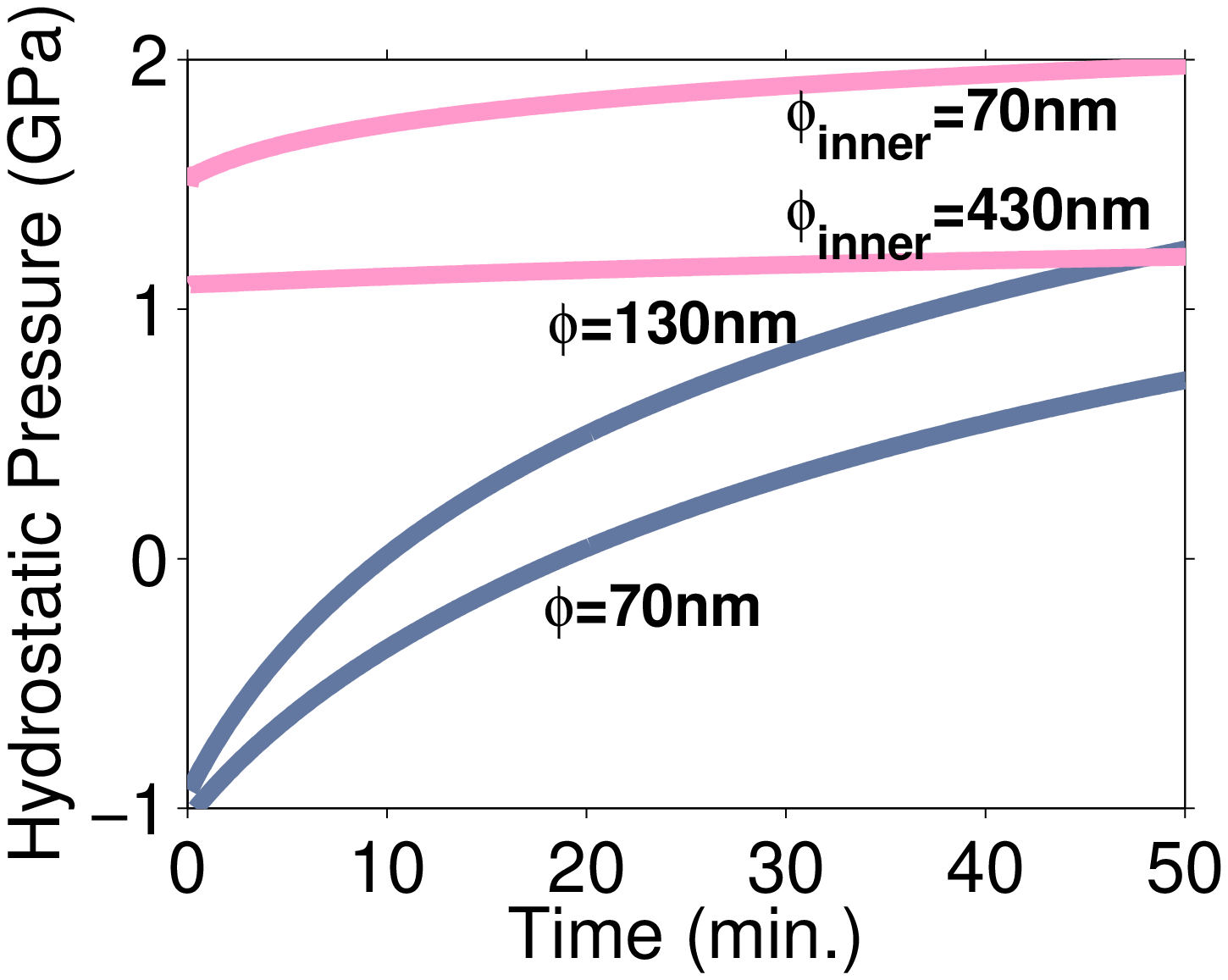}}
  \end{overpic}}  
\caption{\label{fig:fig3}}
\end{figure}

Fig. \ref{fig:fig3}: (a) Modelling of the convex/concave oxidation kinetics using  the plastic model.   b) Evolution of  the theoretical radial stress $\sigma_{r}$ at the Si/SiO$_{2}$ interface. Inset provides the hydrostatic pressure evolution during the oxidation for the different nanostructures.

\begin{figure}
\subfigure[]{\includegraphics[width=6cm]{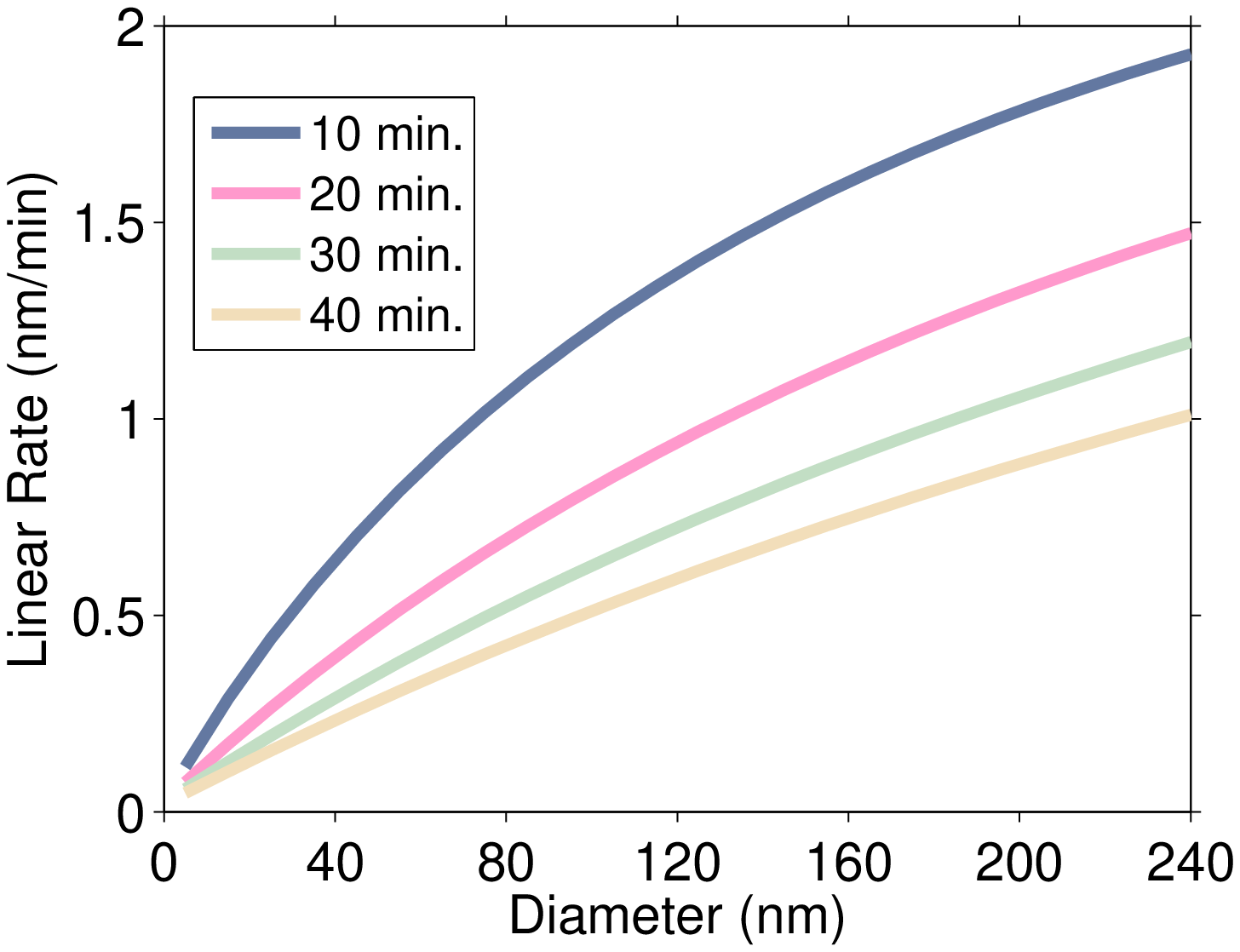}}
\subfigure[]{\includegraphics[width=6cm]{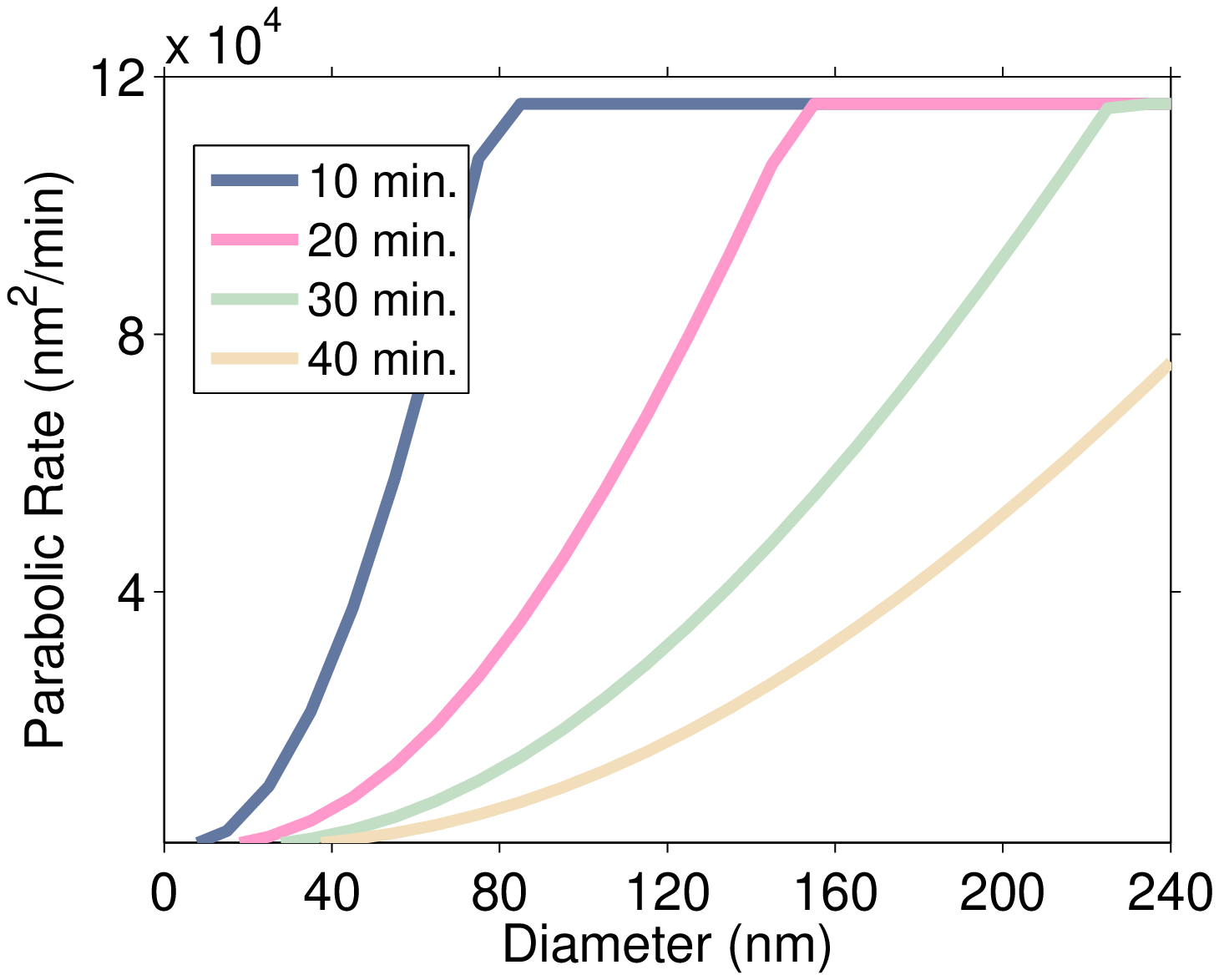}}
 
\subfigure[]{\begin{overpic}[width=6cm]{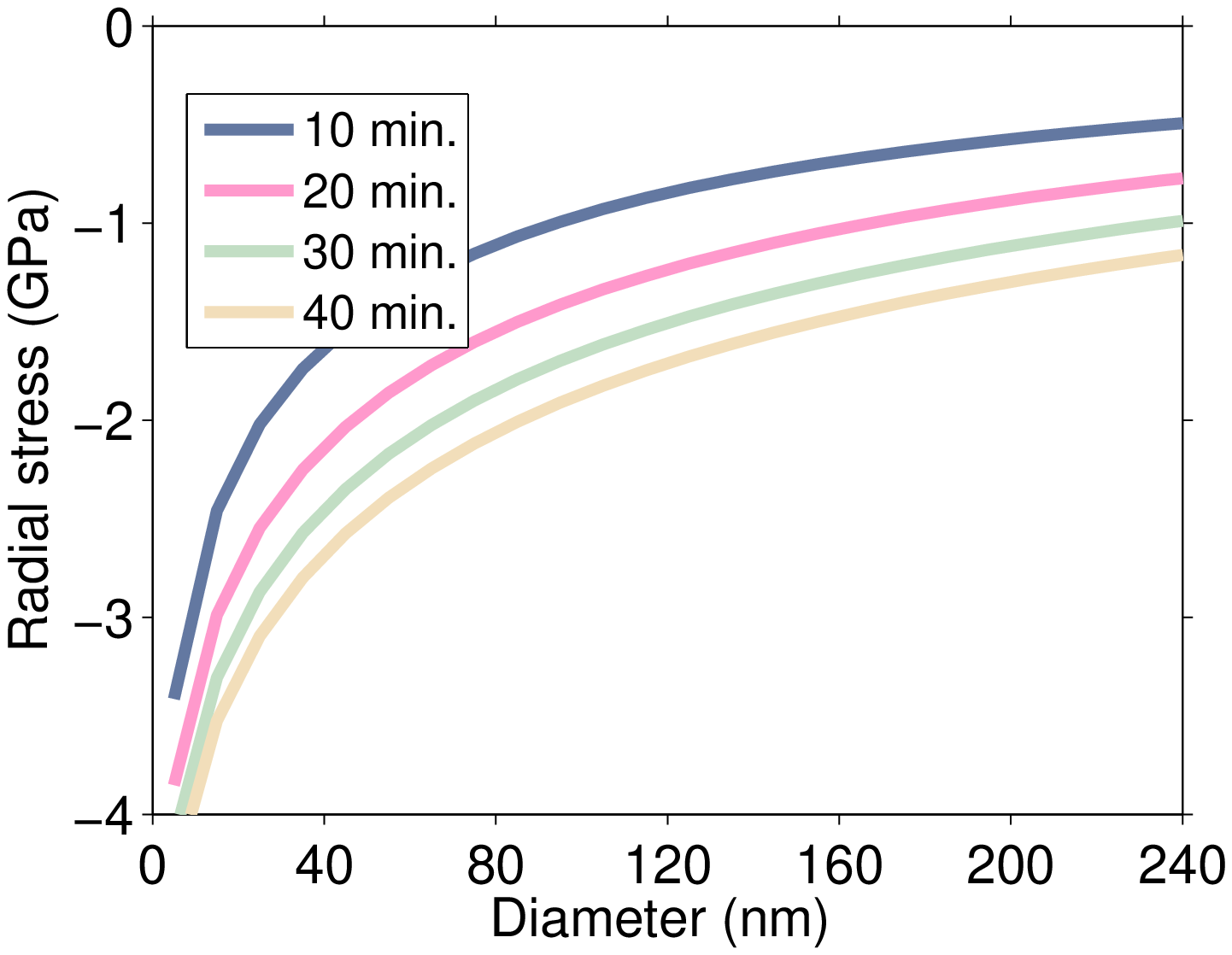}
    \put(45,11){\includegraphics[width=2.7cm]{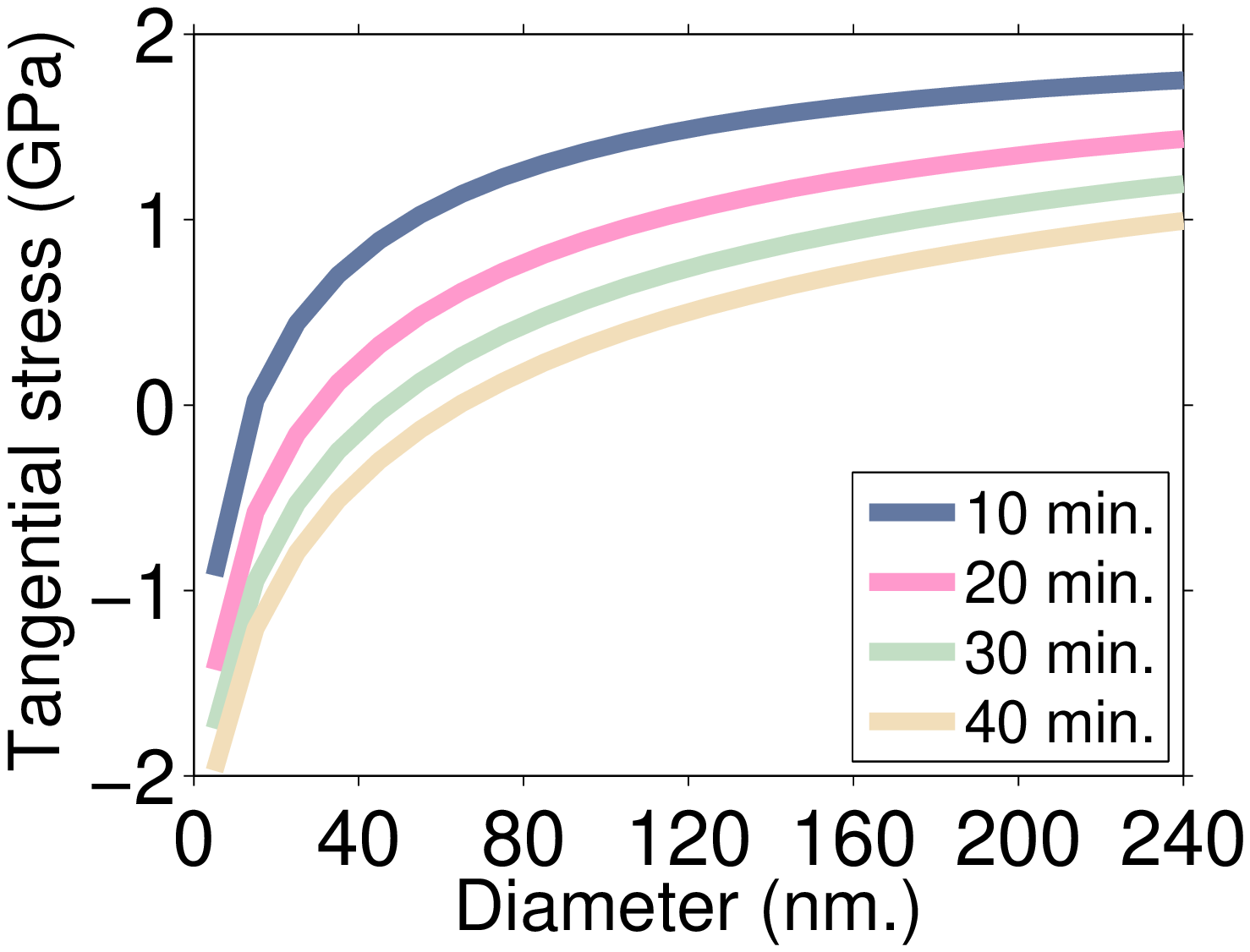}}
              \end{overpic}}
\caption{\label{fig:fig4}}
\end{figure}

Fig. \ref{fig:fig4}:  (a) Linear rate variation with SiNW diameters showing the impact of the radial stress in the initial oxidation regime as the main limiting factor. (b) Highlight of the diffusion limited regime which takes place only when a sufficient oxide thickness has been grown. (c)  Compressive radial stress build-up  with decreasing SiNW diameters. Inset: tangential stress  relaxation.

\newpage
\noindent{FIGURES}

\newpage

\newpage

\newpage

\newpage

\end{document}